\documentclass[
 rsi,
 amsmath,amssymb,
 reprint,%
]{revtex4-1}

\usepackage{graphicx}
\usepackage{dcolumn}
\usepackage{bm}

\usepackage{siunitx}   
\usepackage{color}
\usepackage{textcomp}  
\usepackage{amsmath}
\usepackage{hyphenat}
\usepackage{mathptmx}
\usepackage{float}
\usepackage{placeins}
\usepackage{hyperref}

\makeatletter
\def\@email#1#2{%
 \endgroup
 \patchcmd{\titleblock@produce}
  {\frontmatter@RRAPformat}
  {\frontmatter@RRAPformat{\produce@RRAP{*#1\href{mailto:#2}{#2}}}\frontmatter@RRAPformat}
  {}{}
}%
\makeatother
\begin{document}

\preprint{AIP/123-QED}

\title[title]{Table-top Soft X-ray Source for XAS Experiments with Photon Energies up to 350~eV}
\author{O.~A.~Naranjo-Montoya, M.~Bridger,  R.~Bhar,  L.~Kalkhoff,  M.~Schleberger, H.~Wende, A.~Tarasevitch*, and U.~Bovensiepen}
 \affiliation{
University of Duisburg-Essen, Faculty of Physics and Center for Nanointegration (CENIDE), Lotharstrasse 1, 47057 Duisburg, Germany.
}
\email[]{alexander.tarasevitch@uni-due.de}
\date{\today}

\begin{abstract}
We present a table-top setup for femtosecond x-ray absorption spectroscopy based on high harmonic generation (HHG) in noble gases. Using sub-millijoule pump pulses at a central wavelength of 1550~nm broadband HHG in the range 70 to 350~eV was demonstrated. The HHG coherence lengths of several millimeters were achieved by reaching the nonadiabatic regime of harmonic generation. NEXAFS experiments on the boron K edge of a boron foil and a hexagonal BN (hBN) 2D material demonstrate the capabilities of the setup.
\end{abstract}

\maketitle

\section{\label{sec:Intro}Introduction}

Soft x-ray spectroscopy is a well established methodology which comprises various photon-in photon-out techniques to probe core
level excitations of atoms, molecules, and condensed matter. X-ray absorption spectroscopy
analyzes the transition from a core level to unoccupied states and x-ray emission detects the x-ray fluorescence induced by creation
of a core hole. While both techniques provide information on the excited core levels they complement each other regarding the
valence electronic states. X-ray absorption probes the unoccupied and x-ray emission the occupied valence electronic structure~\cite{degroot01}.

In the spectral range near the resonant core level excitation element and orbital selective insights on the unoccupied valence
electronic structure are obtained primarily in near x-ray absorption fine structure absorption spectroscopy (NEXAFS)~\cite{nexafs:book,degroot01} In the extended spectral range the locally emitted photoelectron wave function upon
resonant core level excitation is scattered back from the neighboring ion cores in molecules and condensed matter, which
leads to interference effects in the absorption cross section detected in extended x-ray absorption fine structure absorption
spectroscopy (EXAFS). This technique analyzes the local ion core structure~\cite{exafs}.

More recently, resonant inelastic x-ray scattering (RIXS) has become available as a tool to measure low energy excitations and
their dispersion relation in momentum space~\cite{rixs,rixs:gelmukhanov}.

Clearly, x-ray spectroscopy is a very important tool in various fields of the natural sciences, like, e.g., material science~\cite{XAS:book,xs:dichroism,xs:magnetic}, interface science~\cite{xas:interface}, coordination chemistry~\cite{xas:chem}, energy conversion and catalysis~\cite{xs:catal}, photosynthesis~\cite{xas:photosyn}, and biology~\cite{xs:biol}. Moreover, it is widely used in conservation of cultural heritage and art~\cite{xas:heritage,xas:art}.

A major effort in setting an x-ray spectrometer is dedicated to the x-ray source. Early efforts exploited the emission of x-ray tubes~\cite{roentgen1919,xray1971}. The required spectral resolution and signal to noise ratio lead to very long times needed for data accumulation in such laboratory based instruments. The high brilliance provided by synchrotron light sources resulted in considerable improvement in this regard and nowadays most x-ray spectroscopy experiments are conducted at synchrotron light sources~\cite{xas:synchrotron}. Higher x-ray photon fluxes and shorter pulse durations became possible after having shifted from storage rings used in synchrotron light sources to linear accelerators in x-ray free
electron lasers which provide dedicated instruments for soft x-ray spectroscopy~\cite{fel2015,fel2022,fel2023,fel:tobias2023}.

The limited access to the necessary beam times has always fueled the development of table-top x-ray light sources and laser-based approaches have led to considerable success. Essentially two mechanisms of light-matter interaction are exploited. For table-top x-ray spectroscopy light sources based on laser plasma generation~\cite{nexafs:plasma,xas:plasma} and on higher harmonic generation (HHG) in gases~\cite{nexafs:popmintchev,xs:ferri} have been developed. Such laser based sources allow generation of ultrashort x-ray pulses~\cite{chang:prl,spielmann:science,joachain00,brabec00,gibson:science,popmintchev:pnas,chen:prl}, which provide highly relevant opportunities for pump-probe experiments and time domain x-ray spectroscopy~\cite{xs:kraus2018}. The plasma based approach provides sufficient photon flux also at hard x-ray photon energies. HHG allows shorter pulse duration~\cite{sidiropoulos:atto} reaching the attosecond limit; a development that was recognized by awarding the Nobel Prize in Physics 2023 to P.~Agostini, A.~L'Huillier, and F.~Krausz.

Using mid-infrared femtosecond pump pulses HHG with photon energies exceeding 1 keV became available~\cite{nexafs:popmintchev,popmintchev:science}, and for competing with the accelerator based sources in terms of the x-ray photon yield a phase matched harmonic emission from a larger atomic ensemble becomes essential. The phase matched HHG by adjusting the gas pressure in the interaction volume has been discussed and demonstrated for HHG in gas jets \cite{constant:prl,geissler00,tempea00,heyl:jpb,tao:njp}, in hollow core waveguides (HCWs) \cite{durfee:prl,chen:prl,popmintchev:pnas,popmintchev:science,chen:pnas,garcia:NJP} or both \cite{meyer:pra}. Using HCWs in longitudinal geometry for the HHG is more promising, because of longer interaction lengths compared to transverse geometry (jets, etc.). Important limitations in this case, are absorption~\cite{constant:prl} and the necessity to keep the gas pressure in the HCW constant along the capillary to provide a long coherence length of the HHG.

For the phase matching two types of the phase contributions have to be taken into account: the time independent (neutral gas dispersion and geometrical contribution)  and time dependent (plasma dispersion and nonlinear refraction). The time dependent one comes into play at higher pump intensities and the HHG switches correspondingly from adiabatic to nonadiabatic \cite{rae:pra,geissler00,brabec00} (or time gated \cite{garcia:NJP}) regime.

In this paper we demonstrate a table-top x-ray source based on HHG in noble gases using low energy (compared to ~\cite{nexafs:popmintchev,popmintchev:science}) pump pulses at the wavelength of 1550~nm. The generated x-ray pulses were applied to the measurements of the x-ray absorption spectra (XAS) on the boron K-edge. We show that switching to the nonadiabatic regime is essential for the phase matched harmonic emission.
\begin{figure*}[t]
\centering
\includegraphics[width=0.75\textwidth]{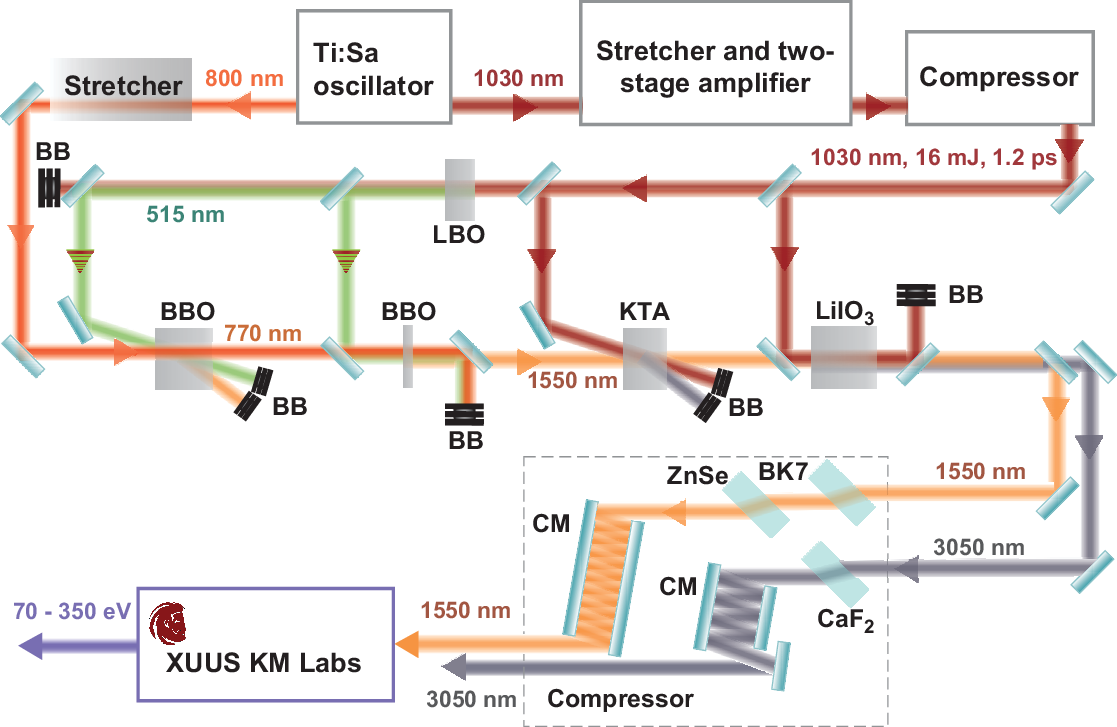}
\caption{Experimental schematic. A broadband Ti:Sa oscillator is used as a seed source for both the OPCPA itself and for the pump channel. The 6~fs pulses at the wavelength of 800~nm are stretched and used as seed for the four stage OPCPA, consisting of two BBO crystals followed by KTA and LiIO$_3$ stages. The pump channel is seeded with 1030~nm pulses and consists of a stretcher, two-stage amplifier and compressor. Part of the pump radiation is converted to the second harmonic (SH) in an LBO crystal for pumping of BBO stages. BB stands for beam blocks. After the amplification the pulses are compressed using CM and MD in ZnSe, BK7, and CaF$_2$. The pulses at the wavelength of 1550~nm are used for pumping of the harmonic source.}
\label{fig1}
\end{figure*}

\section{Experimental set-up}
The pump laser for the HHG was an upgraded version of the optical parametric chirped pulse amplification (OPCPA) set-up reported in~\cite{bridger}. Its schematic is depicted in Fig.~\ref{fig1}. The system was seeded by a Ti:sapphire (Ti:Sa) oscillator producing 6~fs pulses at the central wavelength of 800~nm with an energy of 2.5~nJ per pulse and the repetition rate of 80~MHz. A fraction at the wavelength of 1030$\pm$2~nm of the broad-band oscillator emission was used to seed the OPCPA pump channel. The 1030~nm seed pulses were stretched to a 2~ns duration and amplified in a two-stage amplifier to the energy of 27~mJ. After the amplification, the pulses were compressed to a duration of 1.2~ps. A 1~mJ fraction of the compressed pulse energy was frequency doubled and used to pump the first two OPCPA stages.

The 6~fs seed pulses at the central wavelength of 800~nm provided by the Ti:Sa oscillator were first stretched with the help of a 10~cm BK7 slab to the duration of 3~ps full width at half maximum (FWHM), which is longer than the pump pulse duration. By changing the delay between the pump and the strongly stretched seed pulses, one could choose the central wavelength at which the amplification takes place.

The set-up was running at the repetition rate of 100~Hz. In the first OPCPA stage the pump (wavelength 515~nm) and the seed were interacting noncollinearly in a 3.9~mm thick type~I BBO crystal ($\theta$ = 24.3$^\circ$). The noncollinear interaction with the angle of 2.4$^\circ$($\theta$ = 24.3$^\circ$) allowed reaching high amplification while keeping a broad amplification spectrum. The next OPCPA stage was used for the difference frequency generation. The pump at the wavelength of 515 nm and the signal beam with the central wavelength shifted to 770~nm were interacting collinearly in order to avoid the angular chirp of the idler wave. In order to maintain the broad spectral width, a much shorter (0.6~mm) type~I BBO crystal was used. The idler pulses of the second stage at the central wavelength of 1550~nm were amplified in the third OPCPA stage. A 3~mm thick type~II KTA crystal ($\theta$ = 48.6$^\circ$) was pumped by the pulses at the wavelength of 1030~nm in the noncollinear geometry (noncollinear angle of $4.2 ^\circ$), providing high amplification in a wide spectrum. In the fourth and last OPCPA stage, pumped by the 1030~nm pulses, the radiation at the wavelength of 1550~nm (signal wave) was further amplified up to the energy of 1.8~mJ per pulse. For this purpose a 4.5~mm thick type~I LiIO$_3$ crystal was chosen because of its wide amplification bandwidth for a collinear interaction. The energy of the idler pulses at the wavelength of 3050 nm reached 0.8~mJ.

\begin{figure}[t]
\centering
\includegraphics[width=0.4\textwidth]{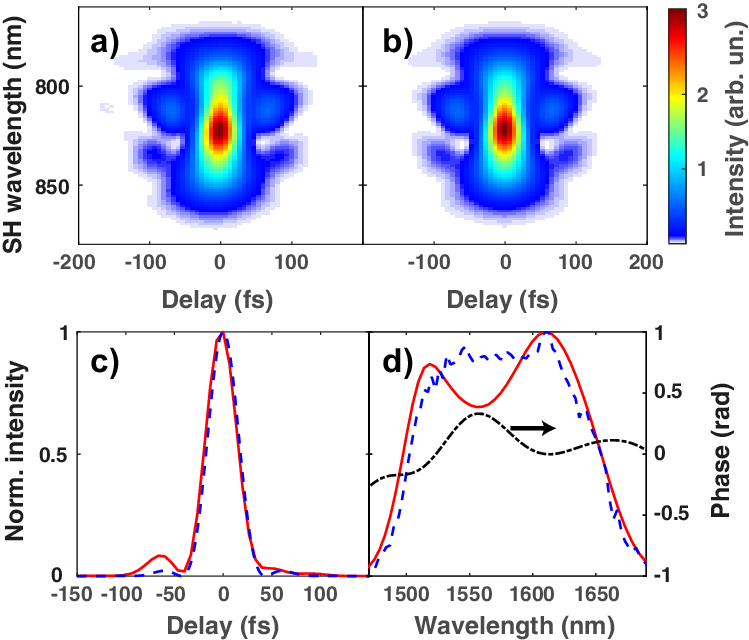}
\caption{Measured (a) and retrieved (b) SH FROG traces of the pulses at the central wavelengths of 1550~nm. The retrieved pulse shape (37~fs
FWHM) and spectrum are shown with red solid lines in the panels (c) and~(d), respectively. The blue dashed lines in (c) and (d) represent correspondingly the bandwidth limited pulse (35~fs FWHM) and the independently measured spectrum. The black dash-dotted line in (d) shows the retrieved spectral phase.}
\label{fig2}
\end{figure}
For the compression of both signal and idler pulses, combinations of material dispersion (MD) and chirped mirror (CM) dispersion were employed~(Fig.~\ref{fig1}). Such a combination allows better compensation of the second and third order dispersion and the possibility of dispersion fine-tuning. The pulse durations were determined using FROG~\cite{trebino}. An example of the measured and retrieved second harmonic FROG traces for the signal pulse is shown in Fig.~\ref{fig2}a and b. The retrieved pulse (Fig.~\ref{fig2}c) had the duration of 37~fs. The retrieved spectrum was in qualitative agreement with the independently measured one (Fig.~\ref{fig2}d). The pulse energy after the compression was 1.6~mJ. The duration of the compressed idler pulses was found to be 43~fs using a third harmonic FROG. The energy of the compressed pulses was 0.6~mJ.

\begin{figure}[b]
\centering
\includegraphics[width=0.48\textwidth]{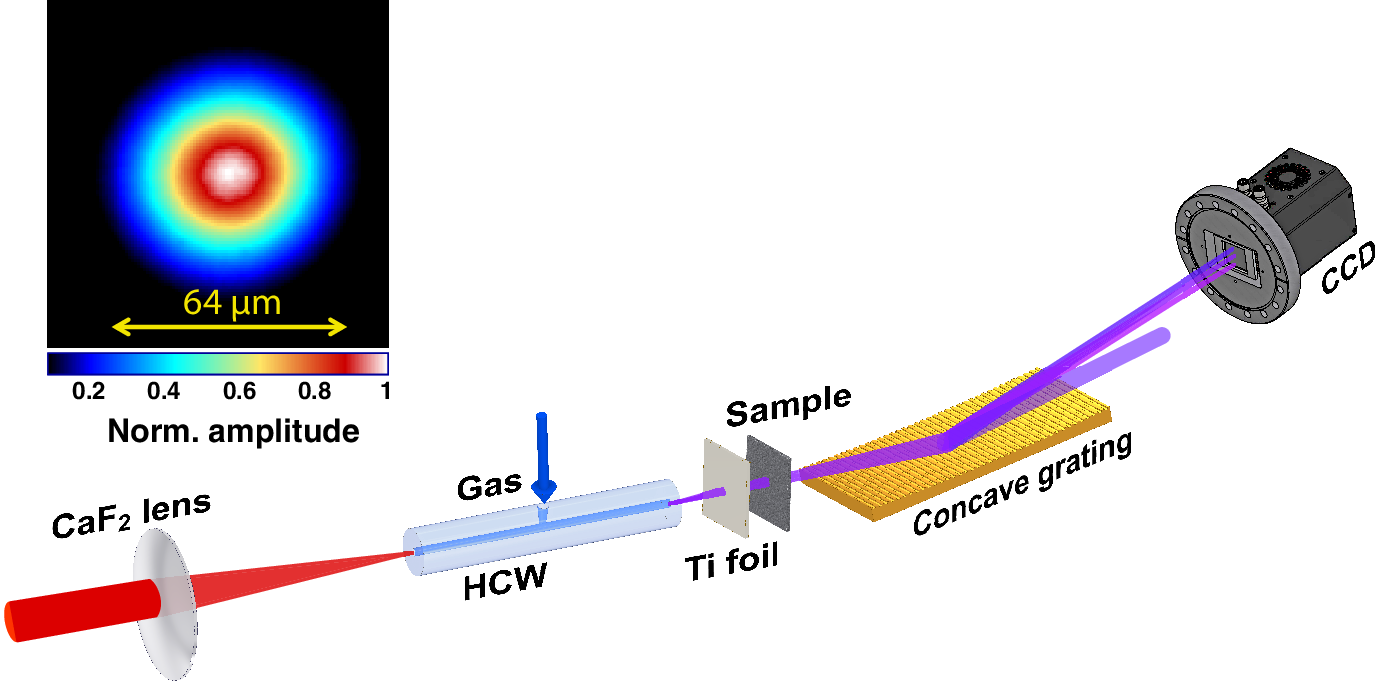}
\caption{The beam at the wavelength of 1550~nm is focused at the entrance of the HCW using a CaF2 lens. The noble gas (Ar, Ne or He) is injected through a small hole in the center of the HCW. The emission at the output of the HCW is filtered with a Ti foil and can be used for the XAS of a sample. The transmitted radiation is spectrally resolved with the help of a concave grating and recorded using a back-side illuminated CCD camera. Inset: energy distribution in the focal plane of the CaF$_2$ lens. }
\label{fig3}
\end{figure}
For the HHG generation a specially adapted "XUUS" source from KM Labs (USA) was used~\cite{xuus}. The signal beam at the wavelength of 1550~nm was coupled into HCW using a 300~mm focal length CaF$_2$ lens~(Fig.~\ref{fig3}). The HCW with an inner diameter of 100~$\mu $m and the length of 10~mm was filled with argon, neon or helium. For the optimum matching of the beam and the EH$_{11}$ mode of the HCW the beam waist diameter at 1/e$^2$ was set to 64~$\mu$m. The inset in Fig.~\ref{fig3} shows the beam energy distribution in the focal plane of the lens. The gas at the pressure up to 30~bar was injected through a hole in the middle of the HCW (Fig.~\ref{fig3}) and  evacuated at its both sides with four roughing pumps.

The emitted radiation from the HCW was filtered with the help of a 200~nm Ti foil and recorded using a McPherson model 251MX flat field spectrometer with a Greateyes (GE 2048 515 B1 UV1) charge-coupled device (CCD) camera~(Fig.~\ref{fig3}). The spectrometer chamber was pumped out with an additional vacuum pump down to the pressure of 10$^{-5}$--10$^{-3}$~mbar depending on the experimental conditions. The spectrometer was equipped with a pair of grazing incidence gratings on concave spherical substrates. The lower photon energy grating (LEG) with 1200~groves/mm and the higher photon energy grating (HEG) with 2400~groves/mm had nominal wavelength ranges of about 5 to 20~nm (250 to 60~eV) and about 1 to 6~nm (1240 to 200~eV).
\begin{figure}[b]
\centering
\includegraphics[width=0.4\textwidth]{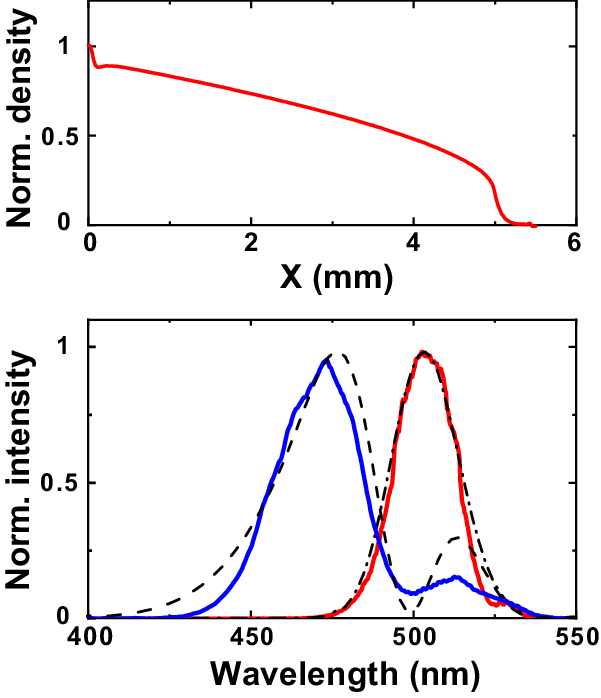}
\caption{Top panel: calculated gas density distribution along the axis of the HCW ($X$-axis). The distribution is symmetric with respect to the middle of the waveguide (X =0). Bottom panel: measured (solid lines) and calculated (dashed and dash-dotted lines) third harmonic spectra at the output of a He-filled HCW. The pump intensity was about $5\times10^{14}$~W/cm$^2$. Red solid and black dash-dotted lines correspond to a low He pressure. Blue solid and black dashed lines represent the spectra for a high He pressure.}
\label{fig4}
\end{figure}

\begin{figure*}
\centering
\includegraphics[width=0.85\textwidth]{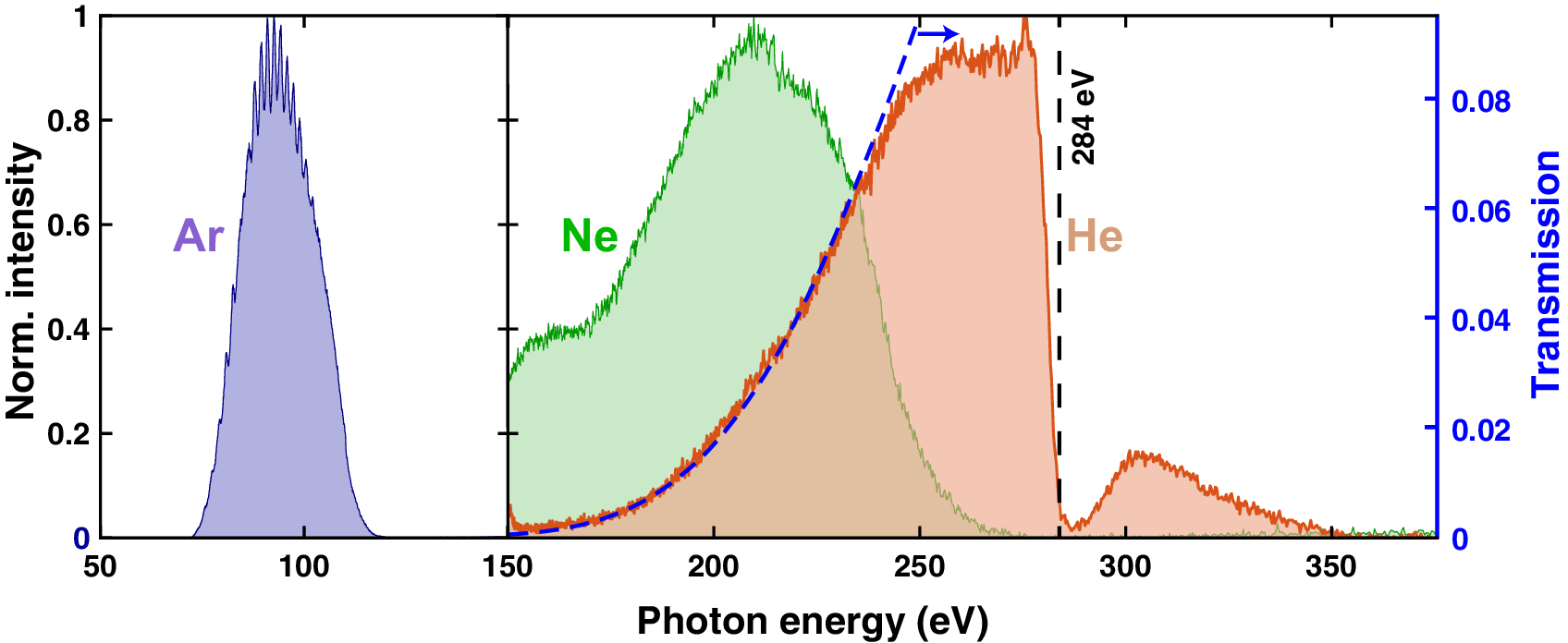}
\caption{Normalized high harmonic emission in Ar (left panel), Ne, and He (right panel) using the 1550~nm fundamental wavelength. In case of Ar the spectrum was recorded using the LEG, while for Ne and He the HEG was used. The pump radiation and lower harmonics were cut using a 200~nm thick Ti-foil. The dashed blue line corresponds to transmission of a 3~mm path in He at 5~bar and the Ti-foil.}
\label{fig5}
\end{figure*}

\section{Switching to the nonadiabatic regime}
Due to a constant gas flow through the HCW, a gas density drop is expected along the waveguide axis. Figure~\ref{fig4}, top panel shows the density distribution along the axis calculated using an Autodesk CFD (Computational Fluid Dynamics) software. Below we estimate the influence of this gradient on reaching the phase matched harmonic emission.

The wave vector mismatch $\Delta k$ between the q-th harmonic and the corresponding polarization wave induced by the pump can be expressed as (see e.g. \cite{popmintchev:pnas})

\begin{equation}
\label{eq1}
\Delta k = - q\frac{2\pi}{\lambda_L}(n_L - 1)(1 - \eta)p + qN_ar_e\lambda_L\eta p + q\frac{u_{nm}^2\lambda_L}{4\pi a^2},
\end{equation}
\noindent where $n_L$ is the refractive index of the gas at the pump wavelength $\lambda_L$, $\eta$ is the ionisation fraction of the gas, $N_a$ is the number density of atoms, $r_e$ is the classical electron radius, $u_{mn}$ is the capillary mode factor, and $a$ is the capillary radius. Both $n_L$ and $N_a$ are taken at the pressure of 1~bar and $p$ is the gas pressure in bar. The factor $1-\eta$ takes into account the reduction of neutral atoms due to ionization. In~(\ref{eq1}) we have neglected the nonlinear refractive index of the gas.

The phase matching pressure tolerance is given by $\Delta p=\Delta k_{FWHM}(d\Delta k/dp)^{-1}$. Above $\Delta k_{FWHM}$ corresponds to the full width at half maximum of harmonic intensity as a function of $\Delta k$ and satisfies the condition $\Delta k_{FWHM}=5.56/L$ \cite{dK}, where $L$ is the interaction (HCW) length. In the adiabatic regime with $\eta\approx0$ one gets:
\begin{equation}
\label{eq2}
\Delta p(\eta=0)=\frac{5.56}{2\pi}\frac{1}{n_L-1}\frac{\lambda_q}{L} ,
\end{equation}

\noindent above $\Delta p(\eta=0)$ is the pressure tolerance in the adiabatic regime and $\lambda_q$ is the wavelength of the q-th harmonic. Simple estimations show that for harmonics with about $1\:keV$ photon energy using $\lambda_L=$1550~nm and $L=$ 1~cm $\Delta p(\eta=0)$ lies in the mbar range and no phase matching along the HCW is possible.

At higher intensities with the developing ionisation the phase matching can be achieved for the fraction of a pump pulse because of the time dependent $\eta$ (nonadiabatic or time gated regime~\cite{rae:pra,geissler00,brabec00,garcia:NJP}). However, the pressure tolerance can be sufficiently increased. Indeed, it follows from~(\ref{eq1}) that
\begin{equation}
\label{eq3}
\Delta p (\eta)=\frac{\Delta p(\eta=0)}{1-\eta/\eta_{cr}},
\end{equation}
\noindent where $\Delta p(\eta)$ is the pressure tolerance in the nonadiabatic regime and $\eta_{cr}$ is the critical ionization fraction, starting from which the phase matching is no more possible ($\eta_{cr}= [\lambda_L^2N_ar_e/(2\pi(n_L-1))+1]^{-1}$~\cite{popmintchev:pnas,durfee:prl}).

The possibility of switching to the nonadiabatic regime in our experiments was checked for helium, because it has the highest ionization potential. The ionisation onset was monitored by measuring the blue shift of the spectrum of the third harmonic at the HCW output using a compact Ocean Optics USB spectrometer (Fig.~\ref{fig4}, bottom panel). The pump intensity was about $5\times10^{14}$~W/cm$^2$. At low He  input pressures  (<100~mbar) the generated electron density is too low and no shift is observed (red solid curve). At 4~bar the blue shift $\Delta\lambda_{3\omega}$ of about 25~nm was measured. Assuming the ratio of the fundamental shift $\Delta\lambda_{\omega}$ to that of the third harmonic to be about 2, as our simulations show, one gets $\Delta\lambda_{\omega}\approx50~nm$. This shift corresponds to the electron production rate of $d\langle n_e\rangle/dt\approx10^{15}$~cm$^{-3}$~fs$^{-1}$, where the angle brackets indicate averaging along the HCW. Dividing this rate over $N_a\langle p\rangle$ one gets $d\eta/dt\approx 2\times10^{-5}\:fs^{-1}$. At this rate the ionization fraction $\eta$ can approach $\eta_{cr}$ (which is about 10$^{-3}$) within the pulse duration, and according to~(\ref{eq3}) a larger pressure tolerance can be achieved.

\section{High Harmonic Generation}
Figure~\ref{fig5} shows the harmonic spectra generated in the nonadiabatic regime using Ar, Ne, and He. Altogether the spectra cover the range from about 70 to about 350~eV. For maximum harmonic emission the pressures and pump energies were 1.5~bar and 0.3~mJ for Ar, 3.7~bar and 0.7~mJ for Ne, and 9.3~bar and 0.8~mJ for He. Furthermore the pulse duration was increased to 54~fs. The photon fluxes at the maxima of the spectra were 10$^7$, 10$^4$, and 10$^5$~ph/s in a fractional bandwidth of 1\% for Ar, Ne, and He, respectively. The high energy cut-offs of the spectra follow the well-known low $\hbar\omega= I_p+3.17U_p$~\cite{corkum93} with the ponderomotive potential $U_p= 9.33\times 10^{14}I_L\lambda^2_L$, where $I_p$ and $I_L$ are the gas ionization potential and laser intensity, respectively. The gap at about 284~eV in the case of He corresponds to the carbon K-edge and is due to the carbon contamination of the spectrometer.

\begin{figure}[t]
\centering
\includegraphics[width=0.45\textwidth]{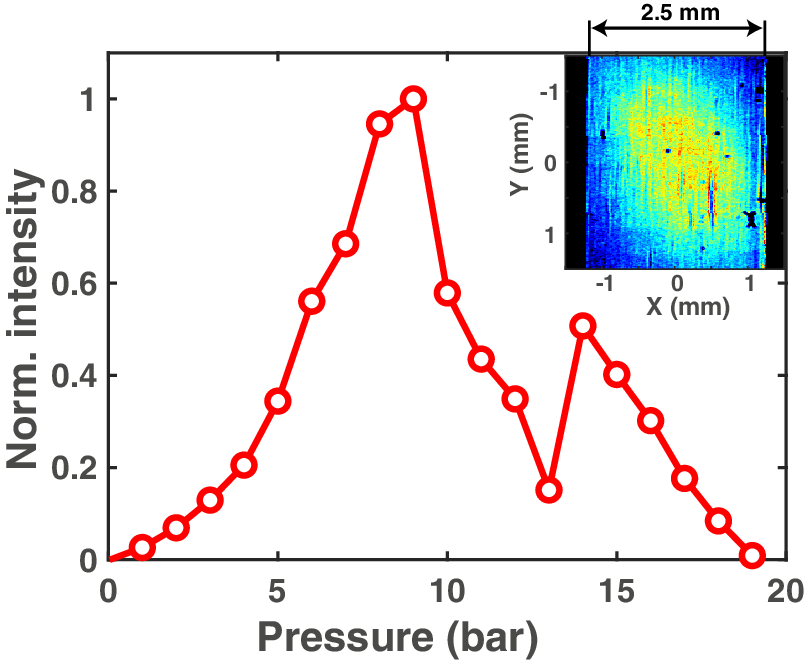}
\caption{Pressure dependence of the HHG intensity in He. Inset: Spatial distribution of the harmonic emission at the position of the entrance slit of the spectrometer}
\label{fig6}
\end{figure}
The decay on the low energy side of the spectra is due to the absorption in the gas and in the Ti foil~\cite{abs, nagasaka20}. The dashed blue line in Fig.~\ref{fig5} shows the calculated transmission of He together with the foil. The He pressure was assumed to be 5~bar, which is the average pressure at the end of the capillary according to Fig.~\ref{fig4}. The path length in He was taken to be 3~mm corresponding to the best match to the measured spectrum. The good agreement indicates, that in spite of a considerable pressure gradient in the HCW (Fig.~\ref{fig4}, left panel) switching to the nonadiabatic regime allows the effective HHG length of several millimeters.

The pressure dependence of the intensity of the harmonic emission for the He-filled HCW is shown in Fig.~\ref{fig6}. The quadratic growth of the intensity and saturation at a certain "optimum" pressure indicates the onset of the phase matched HHG. The shape of the harmonic beam at the entrance slit of the spectrometer can be evaluated from its image (see inset to Fig.~\ref{fig6}) recorded with the CCD-camera using HCW filled with He. The image is taken in the 0$^{th}$ diffraction order of the spectrometer grating with the slit widely open (2.5~mm width). The beam is slightly elliptic with vertical and horizontal sizes of about 2~mm FWHM.

\section{Near Edge X-ray Absorption Fine Structure (NEXAFS)}

The generated harmonic spectra allow XAS measurements on carbon and boron K-edges. In our experiments we have focused on the boron edge, which has been shown to be very informative in several boron compounds~\cite{jimenez}. A set of two samples with different boron configurations was placed in the harmonic beam and the transmitted spectrum was measured with our spectrometer described above (see Fig.~\ref{fig3}). A vacuum filter wheel was used as a sample holder and allowed a rapid exchange of the samples without breaking the vacuum. The first studied sample was a 100 nm thick boron foil, supported by a nickel mesh (Lebow Co). The second one was a 5~mm$\times$5~mm stack of hBN monolayers with a total thickness of 96~nm~\footnote{The stack was prepared following a fabrication protocol described in detail in~\cite{kalkhoff:23}, with the modification of substituting the APS solution for 10~\% nitric acid.}. The sample was supported by a Si$_3$N$_4$ substrate with a thickness of 200 nm.

\begin{figure}[t]
\centering
\includegraphics[width=0.48\textwidth]{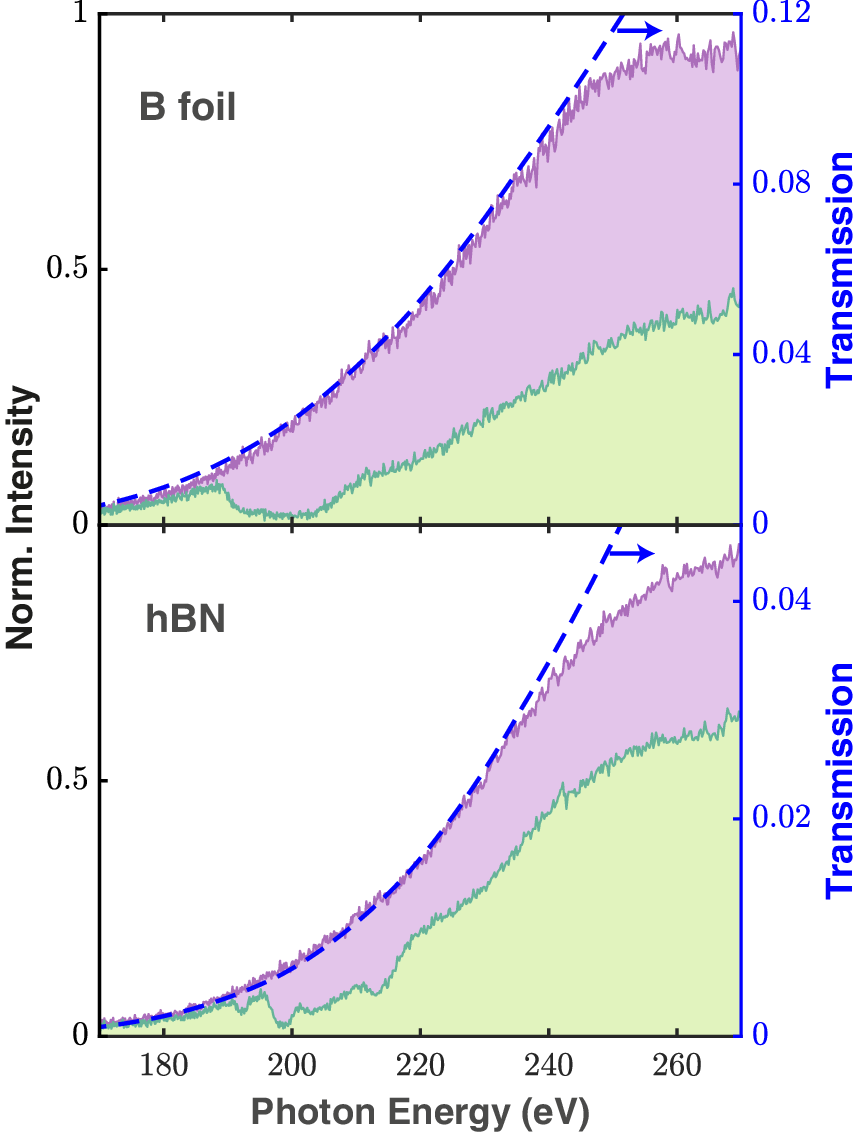}
\caption{Comparison of the transmission spectra (green-filled areas) and reference harmonic spectra (violet-filled area) The samples were a boron foil (top) and an hBN sample (bottom). In the top panel the dashed blue line corresponds to absorption in 200~nm Ti-filter and a path of 3~mm in He at the pressure of 5~bar (top). In the bottom panel additionally a 200~mm Si$_3$N$_4$ substrate is taken into account and the best fit is achieved with a 2.5~mm path. }
\label{fig7}
\end{figure}
Figure~\ref{fig7}, shows the corresponding transmission spectra of the samples (in green) together with the reference HHG spectra (in violet), which were captured without the samples in the beam path. The spectra are normalized to the maxima of the corresponding reference. Each spectrum was obtained in a total time of 20 to 30 min. A clear difference in the transmitted spectra can be observed.

\begin{figure}[t]
\centering
\includegraphics[width=0.45\textwidth]{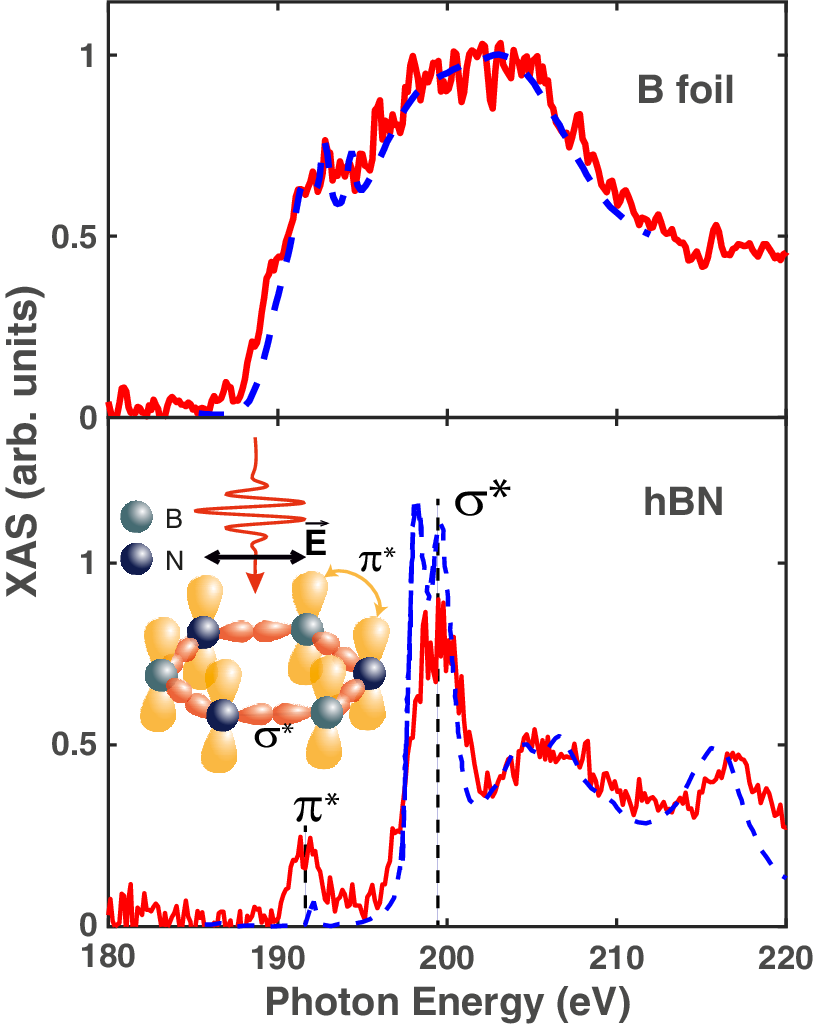}
\caption{XAS for a boron foil (top panel) and a hBN (bottom panel). Red solid lines represent the absorption spectra calculated from those shown in Fig.~\ref{fig7}. The blue dashed lines correspond to the measurements by Jim\'{e}nez et al.~\cite{jimenez} (top panel) and Wang et al.~\cite{wang} (bottom panel). Vertical black dashed lines indicate the positions of the $\pi^*$- and most intense $\sigma^*$ resonance of hBN. Inset in the bottom panel illustrates schematically the orientation of the electric field of the pulse with respect to the $\pi^*$ and $\sigma^*$ orbitals in the 2D hBN layers.}
\label{fig8}
\end{figure}

Figure~\ref{fig8} represents the retrieved XAS for the boron foil and the hBN sample. The retrieved spectrum of the boron foil (Fig.~\ref{fig8}, top panel) is in good agreement with the one recorded by Jim\'{e}nez et al.~\cite{jimenez} using polycrystalline boron at the Stanford Synchrotron Radiation Lightsource. Both spectra show a step starting with the edge jump around 190~eV and going up to around 195~eV, followed by a broad feature of higher amplitude up to~210~eV.

The retrieved NEXAFS of hBN agrees well with the measurements of the XAS fine structures by Wang et al.~\cite{wang}. The well-resolved spectral structure can be attributed to the known resonances of hBN. The spectral resolution of 2~eV was estimated by comparing the measured and the reference curves in the Fig.~\ref{fig8}, bottom.  The peak at 192~eV is the $\pi^*$ resonance, while the features at 198, 199, 204, 206 and 213~eV are due to different $\sigma^*$ resonances~\cite{wang}. The relatively weak contribution due to the $\pi^*$ resonance in our measurements can be attributed (as well as in~\cite{wang}) to the fact that the HHG beam was nearly orthogonal to the sample, which is a 2D material. In this case the electric field, which lies predominantly in the plane of the hBN layers (see inset in Fig.~\ref{fig8}) cannot effectively excite the $\pi^*$ resonance, because for the transition 1s - $\pi^*$ the scalar product of the field and the corresponding matrix element of the dipole momentum is small.

\section{Conclusions}
In conclusion, we have presented a table-top soft x-ray source for XAS experiments with a 100~Hz repetition rate. The source is based on HHG in a noble-gas-filled HCW pumped with pulses at the wavelength of 1550~nm. It produces broadband spectra with the photon energies from 70 to 350~eV. Bcause of  a small diameter HCW (100~$\mu$m) the x-ray emission is generated by pump pulses with  the energies below 1~mJ. In spite of a substantial pressure gradient along the HCW, switching to the nonadiabatic regime allows reaching coherence lengths of the HHG of several millimeters. A longer (1~cm) HCW allows a moderate gas consumption and pump rates of the vacuum pumps. With the He filled HCW photon fluxes of about 10$^5$~ph/s at an energy of 270~eV  in a fractional bandwidth of 1\% are reached. The characteristics and the overall performance of the setup have been demonstrated with NEXAFS measurements on the B K-edge with a 2~eV energy resolution and 20--30~min averaging per recorded spectrum for a boron foil and a hBN sample.

The source provides opportunities for femtosecond time-resolved pump-probe XAS measurements.

The higher photon energy of the HHG source can be achieved by switching to the idler pulses with the central wavelength of 3050~nm (see Fig.~\ref{fig1}. Switching to the nonadiabatic regime also with these pulses will be possible after a planned upgrade of the set-up.

\section{Acknowledgments}
We thank Margaret Murnane and Henry Kapteyn for fruitful discussions.

We acknowledge financial support by the Deutsche Forschungsgemeinschaft through SFB 1242 (project number 278162697, TP A05 and C05) and the Bundesministerium für Bildung und Forschung through project number 05K19PG1.

\section{Author declarations}
\subsection{Conflict of Interests}
The authors have no conflicts to disclose.
\subsection{Author Contributions}
\textbf{O.~Naranjo-Montoya:} Investigation(lead), Data Curation (equal), Resources (equal); \textbf{M.~Bridger:} Resources (equal); \textbf{R.~Bhar:} Resources (equal); \textbf{L.~Kalkhof:} Resources (equal); \textbf{M.~Schleberger} Resources (equal); \textbf{H.~Wende:} Conceptualization (equal), Funding Acquisition (equal), Writing/Review \& Editing (equal); \textbf{A.~Tarasevitch:} Investigation(supporting), Resources (equal), Software (lead),  Data Curation (equal), Methodology (leading), Funding Acquisition (equal), Writing/Original Draft Preparation (equal), Writing/Review \& Editing (equal); \textbf{U.~Bovensiepen:} Supervision(lead), Conceptualization (equal), Funding Acquisition (equal), Writing/Original Draft Preparation (equal), Writing/Review \& Editing (equal). All authors provided critical feedback and helped shape the research, analysis and manuscript.
\section{Availability of Data}
The data that support the findings of this study are available from the corresponding author upon reasonable request.


\begin{thebibliography}{57}
\bibitem [{dK()}]{dK}%
  \BibitemOpen
  \href@noop {} {}\bibinfo {note} {The constant 5.56 comes from the FWHM of the
  function sinc$^2(\Delta kL/2)$, which gives the $\Delta k$-dependence of the
  harmonic intensity.}\BibitemShut {Stop}%
\end{thebibliography}
%
%
\end{document}